\documentclass[a4paper,aps,prl,twocolumn,showpacs,reprint]{revtex4-1}

\setcitestyle{super}
\usepackage[utf8]{inputenc}
\usepackage{graphicx}
\usepackage{dcolumn,subfigure}
\usepackage{amsmath,amssymb,color}
\usepackage{textcomp,array,url} 
\usepackage{multirow,booktabs,siunitx}

\def\Item$#1${\item[-] $\displaystyle#1$ 
\hfill\refstepcounter{equation}(\theequation)}

\usepackage[pdfstartview=FitV,bookmarks=true,
            linktocpage=true,colorlinks=true,
            urlcolor=blue,linkcolor=red,
            citecolor=blue]{hyperref}

\begin{document}

\title{Non contact probing of interfacial stiffnesses between two plates\\ by Zero-Group Velocity Lamb modes}
\author{Sylvain Mezil}      
\author{J\'er\^ome Laurent} 
\author{Daniel Royer}       
\author{Claire Prada}       \email{claire.prada@espci.fr}
\affiliation{Institut Langevin, ESPCI ParisTech, CNRS, 1 rue Jussieu, 75238 Paris Cedex 05, France}

\begin{abstract}
A non contact technique using Zero-Group Velocity (ZGV) Lamb modes is developed to probe the bonding between two solid plates coupled by a thin layer. The layer thickness is assumed to be negligible compared with the plate thickness and the acoustic wavelength. The coupling layer is modeled by a normal and a tangential spring to take into account the normal and shear interfacial stresses. Theoretical ZGV frequencies are determined for a symmetrical bi-layer structure and the effect of the interfacial stiffnesses on the cut-off and ZGV frequencies are evaluated. Experiments are conducted with two glass plates bonded by a drop of water, oil, or salol, leading to a few micrometer thick layer. An evaluation of normal and shear stiffnesses, is obtained using ZGV resonances locally excited and detected with laser ultrasonic techniques.
\end{abstract}

\maketitle


The increasing use of adhesive bonding in the industry has been motivated by the need of stronger and lighter structures. This technique is more suitable for providing continuous adhesion properties and easier to process than other ones like welding, riveting or screwing. The bonding between two solid plates can be probed by various ultrasonic techniques: longitudinal and shear waves reflection or transmission at the interface,\citep{Zarembowitch92,article-chuan} thickness resonances,\cite{Lenoir97} ultrasonic guided waves propagating along the interface.\citep{ article-kazys,article-heller,article-valier-brasier,article-vlasie-acoustical,article-du} All these methods require a modelisation of the ultrasonic wave interaction with the interface supposed homogeneous.\cite{article-rokhlin91} Recently it was shown that Zero-Group Velocity (ZGV) Lamb modes associated with laser ultrasonic techniques, allow a local and non contact measurement of mechanical properties of isotropic or anisotropic plates and shells.\citep{article-prada-laser-based,article-prada-laser-impulse} These ZGV modes, corresponding to a minimum frequency of dispersion curves, also exist in layered plate structures.~\cite{article-nishimiya-relationships} Local ZGV resonances have been used to image the lack of adhesive bond between two plates.\cite{article-prada-laser-ultrasonic}\\

In this letter a non contact method, based on the measurement of ZGV resonance frequencies is proposed to probe interfacial stiffnesses between two plates. An interfacial behavior model~\citep{article-jones-whittier,article-rokhlin91} is used to calculate the dispersion curves and Zero-Group Velocity Lamb modes in a symmetrical structure composed of two plates coupled by a thin layer. The ultrasonic wave interaction with the interface is described by spring boundary conditions. Firstly, the effect of longitudinal and transverse stiffnesses on cut-off and ZGV frequencies is studied. Secondly, experimental results with liquid or solid compliant layers are obtained by laser ultrasonic techniques. Then, the values of spring stiffnesses, estimated from ZGV resonance frequencies, are discussed.\\


\textit{Theoretical model} --- The structure is composed of two identical, isotropic and homogeneous plates with a coupling layer in-between.\citep{ article-vlasie-acoustical} The plate thickness is denoted $h$ and lateral dimensions are supposed infinite. Their mass density is denoted $\rho$ and their bulk wave velocities $c_l$ and $c_t$. The coupling layer thickness $d$ is assumed to be small ($d\ll h$ and $k_ld \ll 1$, where $k_l$ is longitudinal wavenumber) such that its mass effect can be neglected.\cite{article-rokhlin91} Then, the structure is modeled as two plates linked by a normal spring of stiffness $K_n$ and a tangential spring of stiffness $K_t$. In practice, spring stiffnesses per unit area can evolve from 0 (uncoupled plates) up to $10^{17}$~N/m$^{3}$. This upper limit corresponds to a force estimation in a 1D atomic chain model for the layer.\cite{Royer99a}

\begin{figure}[!h]
\centering
\includegraphics[width=0.55\columnwidth]{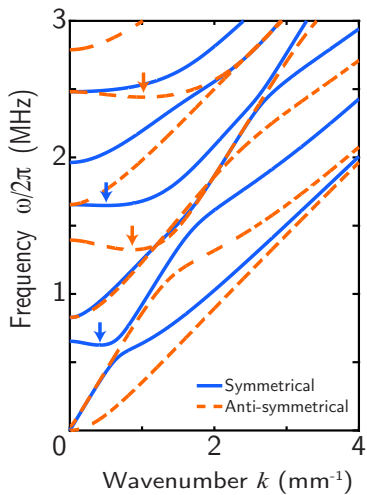}
\caption{Theoretical dispersion curves for two glass plates of thickness $h$=2.08~mm linked by springs of stiffness $K_{n}=3\times 10^{14}$ N/m$^{3}$ and $K_t=10^{11}$ N/m$^{3}$. Arrows indicate ZGV Lamb modes.}
\label{fig:disp}
\end{figure}

Lamb wave propagation is governed by the linear equations of elastodynamics.\cite{Achenbach73} For a given angular frequency $\omega$, the scalar and vector potentials are derived in the whole structure. For $z=\pm h$, the boundary conditions are free tangential and normal stresses. At the interface $z=0$, spring boundary conditions are applied:\cite{article-jones-whittier} the tangential and normal displacement differences between the two plates are equal to the ratios of shear and normal stresses to the spring stiffness $K_t$ and $K_n$, respectively. These analytical developments lead to an equation of the form $[M][B]=[0]$ where $[M]$ is a 8x8 matrix and $[B]$ is a vector composed of the scalar and vector potentials at the interfaces. Non trivial solutions are found when the determinant of $[M]$ vanishes. The propagating modes are defined by their wavenumbers $k$. Then, the corresponding dispersion curves $\omega(k)$ are determined. As the structure is symmetrical, it is possible to separate symmetrical modes (where the in-plane displacements at the bilayer surfaces are equal whereas the normal displacements are opposite) from anti-symmetrical modes (with reversed situation). Finally, the matrix $[M]$ results in two 4x4 matrices. The analytical expressions of corresponding determinants are
\begin{eqnarray}
   S = \Gamma+2 \mu K_n {k_t}^2 p\Big(4 {k}^2 p q\sin[hp]\cos[hq]\nonumber\\
      +\left[{k_t}^2 - 2 {k}^2\right]^2\cos[hp]\sin[hq]\Big),\quad
\label{eq:S}
\end{eqnarray}
for symmetrical ($S$) modes and 
\begin{eqnarray}
   A = \Gamma+2 \mu K_t{k_t}^2 q \Big(\left[{k_t}^2 - 2 {k}^2\right]^2\sin[hp]\cos[hq]\nonumber\\
       +4 {k}^2 p q\cos[hp]\sin[hq]\Big) ,\quad
\label{eq:A}
\end{eqnarray}
for anti-symmetrical ($A$) modes,
with:
\begin{eqnarray}
   \Gamma = -\mu^2\Big[8 p q {k}^2 \left({k_t}^2 - 2 {k}^2\right)^2\left(1-\cos[hp]\cos[hq]\right)\nonumber\\
            +\left(\left[{k_t}^2 - 2 {k}^2\right]^4 + \left[4 {k}^2 p q\right]^2\right)\sin[hp]\sin[hq] \Big],\quad
\label{eq:Gamma}
\end{eqnarray}
where $\mu=\rho {c_t}^2$ is the shear modulus, $k_l=\omega/c_{l}$ and $k_t=\omega/c_{t}$ are the bulk wavenumbers, $p$ and $q$ are the $z$-components of longitudinal and transverse wave vectors, respectively, that satisfy the equations ${k_l}^2=k^2+p^2$ and ${k_t}^2=k^2+q^2$. Due to the symmetry, it can be observed that the symmetrical modes only depend on the longitudinal stiffness $K_n$ while the anti-symmetrical modes only depend on the tangential stiffness $K_t$. This is convenient to study separately the effect of each spring stiffness on the whole system. Fig.~\ref{fig:disp} shows Lamb mode dispersion curves calculated for two identical glass plates of thickness $h=2.08$~mm linked by two springs of stiffnesses $K_n=3\times 10^{14}$ N/m$^{3}$ and $K_t=10^{11}$ N/m$^{3}$. In the frequency range [0-3~MHz] many symmetrical (solid line) and anti-symmetrical (dashed line) modes can be observed.\\

When the wavenumber vanishes, the first two modes decrease to zero frequency, whereas, higher order modes admit a cut-off frequency. Cut-off frequencies are deduced from Eqs.~\ref{eq:S} and~\ref{eq:A} setting $k=0$. It follows
\begin{align}
\frac{\rho \omega^5}{{c_t}^3}\sin\left[\frac{h\omega}{c_t}\right]\left(2K_n\cos\left[\frac{h\omega}     {c_l}\right]-\rho c_l\omega\sin\left[\frac{h\omega}{c_l}\right]\right)=0
\label{eq:CS}
\end{align}
and 
\begin{align}
\frac{\rho \omega^5 c_l}{{c_t}^4}\sin\left[\frac{h\omega}{c_l}\right]\left(2K_t\cos\left[\frac{h\omega}{c_t}\right]-\rho c_t\omega\sin\left[\frac{h\omega}{c_t}\right]\right)=0,
\label{eq:CA}
\end{align}
for symmetrical modes and anti-symmetrical ones, respectively. Eqs~\ref{eq:CS} and~\ref{eq:CA} are the product of three terms leading to different cut-off frequencies:
\begin{itemize}
  \Item    $\omega=0,$\label{eq:FreqC0}
  \Item    $\omega=m \pi c_{l,t}/h \hspace{0.5cm}(m\in\mathbb{N}),$\label{eq:FreqC1}
  \Item    $\omega \tan(h\omega/c_{l,t})=2 K_{n,t}/(\rho c_{l,t}),$\label{eq:FreqC2}
\end{itemize}
for the first, second and last terms respectively. One can observe that cut-off frequencies are of two types: the first two equations (Eqs.~\ref{eq:FreqC0}-\ref{eq:FreqC1}) do not depend on the stiffnesses whereas the solutions of the transcendental equation (Eq.~\ref{eq:FreqC2}) does. Only the last set of cut-off frequencies depends on the bonding between the plates.\\

In the $(\omega,k)$-plane (Fig.~\ref{fig:disp}), Zero-Group Velocity Lamb modes correspond to points where the slope ${d\omega}/{dk}$ of a dispersion curve vanishes whereas the wavenumber $k$ does not.\cite{article-prada-laser-based} The analytical calculation of the derivative over $k$ of Eqs.~\ref{eq:S} and~\ref{eq:A} can directly lead to the ZGV determination, i.e., $\partial S/\partial k$ and $\partial A/\partial k$. For simplicity, the cumbersome equations are not explicitly given here. Unwanted solutions corresponding to bulk waves propagating at velocities $c_l$ ($p=0$) or $c_t$ ($q=0$) are eliminated by dividing Eqs.~\ref{eq:S} and~\ref{eq:A} by the factor $(pq)$. Finally, a symmetric (anti-symmetric) mode have a ZGV point if $S/(pq)$ and $\partial S/\partial k$ simultaneously vanish ($A/(pq)=0$ and $\partial A/\partial k=0$) for a given $(\omega_0, k_0)$ with $k_0\neq0$.\\

\begin{figure}[!ht]
\centering
\subfigure{\includegraphics[width=\columnwidth]{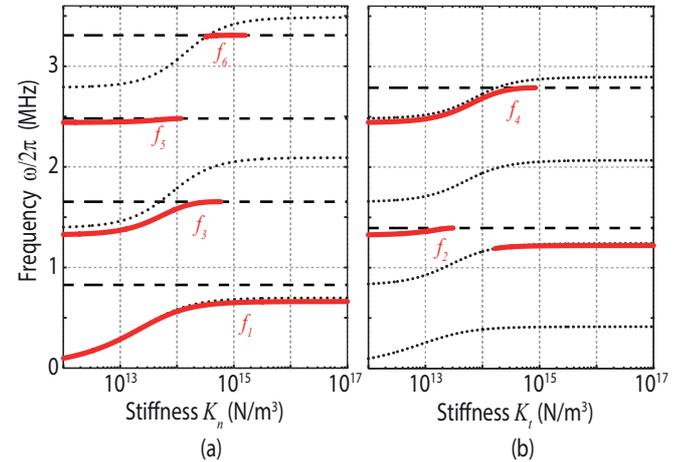}\label{figa:evolZGV_symKL_antKT}}
\subfigure{\label{figb:evolZGV_symKL_antKT}}
\caption{Dependence of cut-off (dashed line and dots) and  ZGV  (solid line) frequencies  on interfacial stiffnesses of (a) symmetric (b) anti-symmetric modes for two glass plates of thickness $h=2.08$~mm.}
\label{fig:evolZGV_symKL_antKT}
\end{figure} 

A numerical study of cut-off and ZGV frequencies as a function of interfacial stiffnesses is conducted for glass plates. Material properties, deduced from experiments,\cite{article-prada-local} are: $h=2.08$~mm, $\rho=2500$~kg.m$^{-3}$, $c_l=5800$~m.s$^{-1}$ and $c_t=3440$~m.s$^{-1}$. From previous considerations, symmetrical and anti-symmetrical modes can be studied independently. Fig.~\ref{fig:evolZGV_symKL_antKT} displays the evolution of cut-off and ZGV frequencies as a function of $K_{n}$ or $K_{t}$, evolving from $10^{12}$ to $10^{17}$~N/m$^{3}$. It appears that a repulsion occurs when two cut-off frequencies, one that depends on interfacial stiffness (Eq.~\ref{eq:FreqC2}) and one that does not (Eq.~\ref{eq:FreqC1}), are equal. This results in a ZGV point of frequency lower than both cut-off frequencies. This effect is similar to the one already established for a single plate: ZGV points result from the repulsion between two modes of the same symmetry having close cut-off frequencies.\cite{Prada08} This repulsion occurs not only when cut-off frequency coincide but also when they remain close enough. The range of stiffness values providing a ZGV mode is not theoretically predicted. It appears from Fig.~\ref{fig:evolZGV_symKL_antKT} that this range varies a lot depending on the modes, and some ZGV modes exist with an important cut-off frequency difference. For the whole range of $K_n$, a low frequency ZGV symmetrical mode can be observed. Its frequency $f_1$ is very sensitive to the value of $K_n$ and is lower than any ZGV frequency of the single plate ($1.323$~MHz for $h=2.08$~mm). Thus, the existence of this low frequency ZGV mode is characteristic of a mechanical coupling between the two plates. From Fig.~\ref{fig:evolZGV_symKL_antKT} it also appears that ZGV frequencies are increasing monotonous functions of $K_n$ or $K_t$. Hence, to one pair of $K_n$, $K_t$ corresponds a unique set of ZGV frequencies. Reciprocally, if the ZGV resonances are experimentally observed, their frequencies directly provide the stiffness values.\\

\begin{figure}[!h]
\centering
\includegraphics[width=0.5\columnwidth]{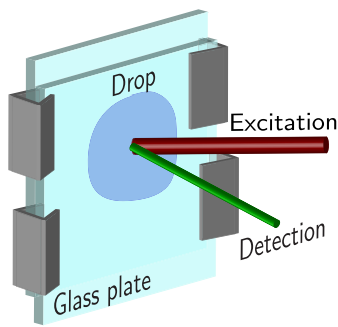}
\caption{Scheme of the experimental set-up.}
\label{fig:setup}
\end{figure}

\textit{Experiments} --- Two glass plates were bonded by a thin liquid or solid layer. ZGV resonance were locally excited and detected by laser ultrasonic techniques. Values of longitudinal and transverse stiffnesses are deduced from the comparison of experimental spectra and theoretical predictions. The experimental set-up is all-optical (Fig.~\ref{fig:setup}). Lamb waves are excited by a Q-switched Nd:YAG laser (1064~nm) with a 20-ns pulse of 8-mJ energy. The 5-mm diameter beam corresponding roughly to the thickness of the structure was chosen to enhance the ZGV mode generation.\cite{article-prada-local} The $10 \times 10$~cm$^2$ glass plates, which properties are given above, are coupled with a liquid (water, oil) or a solid (salol) layer. The salol is liquefied by heating and a drop is deposited on the plates as for the liquids. The drop volume (30~\textmu L) is controlled with a micropipette.\\

The layer thickness, estimated from the ratio of this quantity with the measured area, is within a few micrometers: 6.2 \textmu m for water, 10.3 \textmu m for oil and 5.4~\textmu m for salol. This feature is very small compare to the plate thickness (2.08~mm), thus the bi-layer approximation is valid. A thin layer of aluminium ($\sim 150$~nm) is deposited on one of the glass plates to enhance laser excitation and detection. The normal surface displacement is detected by a 532-nm wavelength heterodyne interferometer in the laser excitation spot. The plates are attached by two bulldog clips on each lateral side and to avoid unwanted constraints on the coupling layer, the plates are shifted vertically, and the lower one is fixed at its bottom (Fig.~\ref{fig:setup}). A measurement without liquid, identical to the measurement with a single plate, ensures that the bulldog clips do not cause a coupling between the plates. Two different liquids and a solid have been studied: distilled water, silicone oil (Rhodorsil 47V50 of density $\rho_{oil}=959$~kg/m$^3$ and viscosity $\eta_{oil}=50.2$~cP) and salol (or phenyl salicylate). The experiments were achieved 5 times for each layer. For repeatability, the plates were carefully cleaned with ethanol then dried. The different spectra are presented in Fig.~\ref{figa:comparaison_exp_theo} and the first ZGV frequencies, as well as their variability, are reported in Tab.~\ref{tab1:resultats}.\\

{\setlength{\extrarowheight}{1.2pt}
\begin{table}[!ht]
\centering
\caption{\label{tab1:resultats}ZGV frequencies measured for different structures ($f<$3~MHz). The frequency uncertainties correspond to repeated layer realizations.}
\begin{tabular}{c  D{,}{\pm}{-1} D{,}{\pm}{-1} D{,}{\pm}{-1} D{,}{\pm}{-1}}
\toprule
\hline\hline
Frequency  & \multicolumn{3}{c}{Bonding layer}             
           & \multicolumn{1}{c}{Single}\\
\cline{2-4}
(kHz)      & \multicolumn{1}{c}{\textbf{Water}} 
           & \multicolumn{1}{c}{\textbf{Oil}} 
           & \multicolumn{1}{c}{\textbf{Salol}} 
           & \multicolumn{1}{c}{plate} \\
\hline
\midrule
	$f_1$  &  629,8    &  548,4    & 655,1  &             \\
	$f_2$  &  1319,1   &  1318,0.1 & 1211,5 &   1320,0.1  \\
	$f_3$  &  1646,4   &  1562,6   &        &             \\
	$f_4$  &  2435,0.1 &  2435,0.1 & 2778,6 &   2436,1    \\
	$f_5$  &           &  2473,1   &        &             \\
\bottomrule
\hline\hline
\end{tabular}
\end{table}

Several ZGV resonances having quality factors $Q>150$ are observed, allowing frequency measurements with 0.1 \% accuracy. The lower resonance frequency $f_1$ corresponding to a ZGV symmetrical mode depends on the coupling layer. It demonstrates a coupling with different bonding forces. Interfacial stiffnesses are estimated by fitting theoretical and experimental ZGV frequencies (Fig.~\ref{fig:comparaison_exp_theo}). From the $f_1$ peak, it follows: $K_n\in[1.43;1.95]\times 10^{15}$N/m$^{3}$ for salol, $[2.8;4.4]\times 10^{14}$N/m$^{3}$ for water, and $[8.0;8.7]\times 10^{13}$N/m$^{3}$ for oil. For these three $K_n$ values, the theory predicts 0, 1 or 2 other resonances associated with symmetrical modes between 1 and 3~MHz: $f_3$ and $f_5$ (Fig.~\ref{fig:evolZGV_symKL_antKT}). The experimental ZGV frequencies given in Tab.~\ref{tab1:resultats} fit very well with theory. For example, the $f_3$ frequency values lead to $K_n\in[2.4;3.2]\times 10^{14}$N/m$^{3}$ for water, agreeing the previous interval found for $f_1$. Thus, the most likely interval for $K_n$ is $[2.8;3.2]\times 10^{14}$N/m$^3$. It is interesting to observe that the peak associated with $f_5$ is theoretically absent for $K_n>1.2 \times 10^{14}$N/m$^{3}$, such as the one associated with $f_3$ for $K_n>6 \times 10^{14}$N/m$^{3}$, which also confirms our predictions. The intervals are clearly distinct for the various coupling layers.\\

\begin{figure*}[!ht]
\centering
  \begin{minipage}[c]{0.70\textwidth}
      \subfigure{\includegraphics[width=\textwidth]{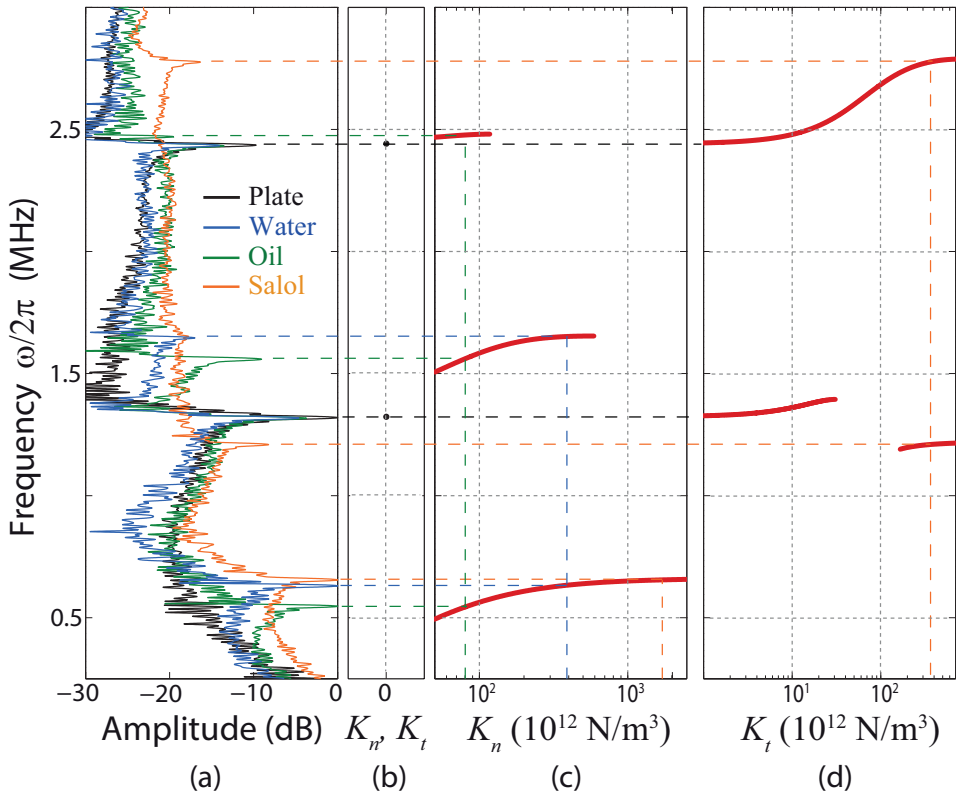}\label{figa:comparaison_exp_theo}}
      \subfigure{\label{figb:comparaison_exp_theo}}
      \subfigure{\label{figc:comparaison_exp_theo}}
      \subfigure{\label{figd:comparaison_exp_theo}}
  \end{minipage}\hfill
  \begin{minipage}[c]{0.27\textwidth}
     \caption{(color online) (a) Experimental spectra for the three different bonding layers, compared with theoretical ZGV frequencies for (b) a single plate, (c) symmetrical modes, $K_n\in[50;3000]\times 10^{12}$N/m$^{3}$, (d) anti-symmetrical modes, $K_t\in[1;700]\times 10^{12}$N/m$^{3}$.}
     \label{fig:comparaison_exp_theo}
  \end{minipage}
\end{figure*}

For both liquids, the second and fourth ZGV peak, $f_2$ and $f_4$ are similar to the single plate resonance frequencies (Tab.~\ref{tab1:resultats}). This corresponds theoretically to a low value of $K_n$ or a low value of $K_t$ (Fig.~\ref{fig:evolZGV_symKL_antKT}). Thus, the previous estimations of $K_n$ imply a weak shear stiffness ($K_t\leqslant $~$10^{11}$N/m$^{3}$) which is reasonable for liquids. The method is not sensitive enough to estimate stiffnesses below this limit. As expected from bulk wave velocities, the mechanical coupling of water is higher than the one of oil. The salol exhibits two different peaks ($f_2$ and $f_4$) that can be linked to $K_t$ values. In this case, it leads to $K_t\in[3.2;4.4] \times 10^{14}$N/m$^{3}$. The normal stiffness is about 4 times higher than the shear stiffness (Tab.~\ref{tab2:resultats}). The ZGV frequencies below 3~MHz were presented, actually the spectra exhibit higher frequency peaks that also agree with the theory.\\ 

\begin{table}[!ht]
\centering
\caption{\label{tab2:resultats}Experimental interface stiffnesses.}
\begin{ruledtabular}
\begin{tabular}{lrrr}
\multicolumn{1}{c}{Bonding}  & \multicolumn{1}{c}{$K_n$}             
                             & \multicolumn{1}{c}{$K_t$}            
                             & \multicolumn{1}{c}{d}                 \\
\multicolumn{1}{c}{Layer}    & \multicolumn{1}{c}{$(10^{14}~N/m^3)$} 
                             & \multicolumn{1}{c}{$(10^{14}~N/m^3)$} 
                             & \multicolumn{1}{c}{($\mu m$)}          \\
\hline
\textbf{Water} & 3.00  $\pm$ 0.20  &   $<$0.001       & 6.20   \\
\textbf{Oil}   & 0.84  $\pm$ 0.04  &   $<$0.001       & 10.30  \\
\textbf{Salol} & 16.90 $\pm$ 2.60  &   3.80 $\pm$ 0.6 & 5.40   \\
\end{tabular}
\end{ruledtabular}
\end{table}
%

In conclusion, an analytic model was applied to calculated Lamb modes guided in a bi-layer structure. The dependence of cut-off and ZGV frequencies as a function of interfacial stiffnesses has been studied. It was demonstrated that symmetrical ZGV frequencies depend on the normal stiffness while anti-symmetrical ZGV frequencies depend on the tangential stiffness. A one to one correspondence between ZGV frequencies and spring stiffnesses was observed so that the knowledge of the ZGV frequencies provides the stiffness values and reciprocally. Experimental evidence of the elastic coupling between two glass plates was achieved with water, oil or salol as a bonding layer. In all cases, the method is sensitive to the normal stiffness. For liquids, the shear interfacial stiffness is too low to be determined. However, the possibility to estimate this parameter is demonstrated for the solid layer. In a further investigation, a three layer model should be applied to take into account the influence of the coupling layer thickness on ZGV resonance frequencies. This method could be also used to study properties of ultra thin layers.\cite{Leopoldes13}\\

The authors thank M. Rousseau, T. Valier-Brasier, and J.-L. Izbicki for fruitful discussions and S. Raetz for helping in the development of the numerical program. This work was supported by the Chaire SAFRAN - ESPCI and the LABEX WIFI (Laboratory of Excellence ANR-10-LABX-24) within the French Program Investments for the Future under reference ANR-10-IDEX-0001-02 PSL*.



\end{document}